\documentclass{WileyMSP-template}
\usepackage{titlesec}
\usepackage[super,sort&compress,comma,square]{natbib}
\usepackage[colorlinks=true,allcolors=blue]{hyperref}
\usepackage{amsmath}
\usepackage{amssymb}

\begin{document}


\title{Experimental Joint Estimation of Phase and Phase Diffusion via  Deterministic Bell Measurements}

\maketitle


\author{Ben Wang*,}
\author{Minghao Mi,}
\author{Huangqiuchen Wang,}
\author{Qian Xie,}
\author{Lijian Zhang*}


\dedication{}

\begin{affiliations}
B. Wang, M. Mi, H. Wang, Q. Xie, L. Zhang\\
National Laboratory of Solid State Microstructures,\\
Key Laboratory of Intelligent Optical Sensing and Manipulation,\\
College of Engineering and Applied Sciences,\\
Jiangsu Physical Science Research Center,\\
Collaborative Innovation Center of Advanced Microstructures,\\
Nanjing University, Nanjing 210093, China\\
E-mail: ben.wang@nju.edu.cn; lijian.zhang@nju.edu.cn
\end{affiliations}



\begin{abstract}
Accurate phase estimation plays a pivotal role in quantum metrology, yet its precision is significantly affected by noise, particularly phase-diffusive noise caused by phase drift. To address this challenge, the joint estimation of phase and phase diffusion has emerged as an effective approach, transforming the problem into a multi-parameter estimation task. However, the incompatibility between optimal measurements for different parameters prevents single-copy measurements from reaching the fundamental precision limits defined by the quantum Cramér–Rao bound. Meanwhile, collective measurements performed on multiple identical copies can mitigate this incompatibility and thus enhance the precision of joint parameter estimation. This work experimentally demonstrates joint phase and phase-diffusion estimation using deterministic Bell measurements on a two-qubit system. A linear optical network is employed to implement both parameter encoding and deterministic Bell measurements, achieving improved estimation precision compared to any separable measurement strategy. This work proposes a new framework for phase estimation under phase-diffusive noise and underscores the substantial advantages of collective measurements in multi-parameter quantum metrology.

\end{abstract}


\section{Introduction}
Quantum metrology leverages quantum resources, such as entanglement and squeezing, to surpass the precision limits inherent in classical measurement techniques.\cite{PhysRevD.23.1693,PhysRevLett.59.278,PhysRevLett.59.2153,doi:10.1126/science.1104149,Higgins2007,Higgins_2009,Daryanoosh2018} Within this framework, accurate phase estimation is pivotal, as many physical observables  can be mapped into phase shifts. For instance, applications such as gravitational wave detection,\cite{Goda2008,Schnabel2010,PhysRevLett.110.181101} high-resolution lithography~\cite{PhysRevLett.85.2733,PhysRevLett.87.013602,Kawabe:07} and quantum imaging~\cite{Moreau2019,PhysRevLett.88.203601,Ono2013} depend critically on the detection of subtle phase variations. Consequently, precise phase estimation serves as a central aspect for advanced metrological protocols.

In realistic scenarios, the inevitable noise in the interferometer will adversely impacts the estimation precision. 
Many types of noise can be pre-calibrated. However, noise arising from the estimated phase itself, such as random phase drift, is difficult to calibrate. This type of noise, known as phase-diffusive noise, causes decoherence, which may reduce or even completely eliminate the quantum advantages that enable enhanced measurement precision.\cite{demkowicz2012elusive,PhysRevA.83.063836,PhysRevA.80.013825,PhysRevLett.102.040403,kacprowicz2010experimental} Although numerous studies have investigated the fundamental precision of phase estimation with known magnitude of phase diffusion,\cite{PhysRevA.81.012305,PhysRevLett.106.153603,PhysRevA.85.043817,PhysRevLett.109.190404,knysh2013estimation} many physical processes involve entirely unknown phase diffusion. Treating the phase and amplitude of phase diffusion as estimable parameters offers a novel and robust solution to this challenge, and related studies have attracted significant attention.
\cite{Vidrighin2014,PhysRevA.92.032114,Jayakumar_2024}
Vidrighin et al. first proposed the a quantum model of this question, and experimentally demonstrated the fundamental precision of joint estimation of phase and phase diffusion at single-copy state level.~\cite{Vidrighin2014}

Moreover, in multi-parameter estimation scenarios, the optimal measurements for different parameters are often incompatible, resulting in inevitable trade-offs in precision.\cite{OZAWA2004367,PhysRevA.94.052108,PhysRevA.105.062442} These trade-offs establish fundamental limitations on the overall precision achievable in multi-parameter estimation and have motivated the formulation of various precision bounds.
\cite{Helstrom1969,holevo2011probabilistic,PhysRevLett.126.120503,Xiang2011,doi:10.1126/science.1138007,Yuan2017,Daryanoosh2018} Such trade-offs limit the overall precision in the joint estimation of phase and phase diffusion. To address this limitation, various studies have utilized collective measurements on multiple copies of quantum states to enhance estimation precision. \cite{Hou2018,Conlon2023}
Notably, theoretical studies have demonstrated that performing Bell measurements—a type of collective measurement—on two copies of quantum states can improve the accuracy of simultaneous estimation of phase and phase diffusion amplitude.~\cite{Vidrighin2014} However, the experimental implementation of Bell measurements remains technically challenging, as existing approaches either depend on probabilistic protocols or require the use of hyper-entangled photon sources combined with complex optical configurations.\cite{PhysRevLett.96.190501,PhysRevLett.118.050501,Barreiro2008} For example, Roccia et al. implemented incomplete and non-deterministic Bell measurements using two-photon interference and subsequently performed quantum tomography to characterize the resulting measurements, thereby inferring the quantum metrological advantage, which is a proof-of-principle demonstration and does not accomplish the practical estimation task.~\cite{Roccia_2018} 
The scheme based on a one-dimensional quantum random walks is a promising method to realize the general positive operator-valued measure (POVM) on photonics system.~\cite{PhysRevLett.114.203602,Hou2018,PhysRevA.100.042302}
Developing a practical optical network capable of performing deterministic Bell measurements via photonic quantum walks on two-copy quantum states that encode phase information under phase-diffusive noise remains an unresolved challenge in the field.

In this work, we experimentally demonstrate that collective measurements on multi-copy systems can enhance the precision of multi-parameter estimation for mixed states. Specifically, we utilize a linear-optical network to prepare two-copy quantum states that encode phase information under phase-diffusive noise and implement deterministic Bell measurements. Our collective measurement scheme achieves approximately an approximate 50\% improvement in estimation precision compared to separable measurements, approaching the ultimate theoretical limit for the two-copy system. Moreover, by adopting the Lu–Wang multi-parameter quantum estimation bound as a benchmark \cite{PhysRevLett.126.120503}, we rigorously validate the superiority of our measurement strategy. These results effectively connect theoretical performance bounds with experimental realizations, offering insights for developing more robust quantum metrology protocols in noisy environments.

\section{Theoretical framework}
In practical interferometric setups, phase-diffusive noise arises as a nondissipative dephasing process in the interferometer arms. This noise can be modeled by applying random phase shifts drawn from a Gaussian distribution with mean $\phi$ and variance $2\Delta^2$, effectively encoding the phase $\phi$ and the phase diffusion $\Delta$ into the quantum state (as illustrated in \textbf{Figure}~\ref{schematic}).
Specifically, for an initial state $\hat{\rho}(0)$, the evolved state under Gaussian phase noise is given by
\begin{equation}
\hat{\rho}_{\phi,\Delta} = \int_{-\infty}^{\infty} d\tilde{\phi}\, \frac{\exp\left(-(\tilde{\phi} - \phi)^2 / (4\Delta^2)\right)}{\sqrt{4\pi \Delta^2}} \, \hat{U}_{\tilde{\phi}} \, \hat{\rho}(0) \, \hat{U}_{\tilde{\phi}}^\dagger,
\end{equation}
where $\hat{U}_{\tilde{\phi}} = \exp(-i \tilde{\phi} \, \hat{a}^\dagger \hat{a})$ implements the phase shift, with $\hat{a}\ (\hat{a}^\dagger)$ denoting the annihilation (creation) operator.\cite{Vidrighin2014,PhysRevA.81.012305} In the Fock basis, the initial quantum state can be written by $\hat{\rho}(0)=\sum_{mn}\rho_{nm}|n\rangle\langle m|$, and the integral yields
\begin{equation}
\hat{\rho}_{\phi,\Delta} = \sum_{n,m} \exp(-\Delta^2 (n-m)^2 + i \phi (n-m)) \, \rho_{nm} \, |n\rangle\langle m|,
\end{equation}
revealing that the diagonal elements remain invariant, thus conserving energy, whereas the off-diagonal elements acquire exponential damping factor $e^{-\Delta^2 (n-m)^2}$ and phase factor $e^{i \phi (n-m)}$.
This encoding process is equivalent to solving the corresponding quantum master equation for phase diffusion.\cite{PhysRevLett.106.153603,PhysRevA.81.012305,PhysRevA.85.043817} For an initial pure qubit state $|\psi_0\rangle = \cos(\theta/2) |0\rangle + \sin(\theta/2) |1\rangle$, in which case the phase-shift operator can be simplified as $\hat{U}_{\tilde{\phi}}=\exp(i\tilde{\phi}\hat{\sigma}_z/2)$ with Pauli operator $\hat{\sigma}_z=\mathrm{diag}[1,-1]$, the density matrix with phase and phase diffusion is
\begin{equation}
\hat{\rho}_{\phi,\Delta} = \begin{pmatrix}
\cos^2\frac{\theta}{2} & \cos\frac{\theta}{2}\sin\frac{\theta}{2} e^{-i\phi - \Delta^2} \\
\cos\frac{\theta}{2}\sin\frac{\theta}{2} e^{i\phi - \Delta^2} & \sin^2\frac{\theta}{2}
\end{pmatrix}.
\end{equation}

To extract information about the parameters $\boldsymbol{x}=(\phi, \Delta)^\top\in\mathbb{R}^2 $, we perform measurements described by the positive operator-valued measure (POVM) $\{\hat{E}_k | \hat{E}_k \ge 0, \sum_k \hat{E}_k = \mathbb{I} \}$. The probability of obtaining the outcome $ k $ is then given by $ p(k|\boldsymbol{x}) = \operatorname{Tr}(\hat{\rho}_{\boldsymbol{x}} \hat{E}_k)$.  The outcomes of such a measurement can be used in a function called the estimator $\check{\boldsymbol{x}}(k)$, which provides an unbiased estimate of $\boldsymbol{x}$. Its precision is quantified by the covariance matrix $\boldsymbol{V}_{\boldsymbol{x}} = \sum_{k} (\check{\boldsymbol{x}}(k) - \boldsymbol{x})(\check{\boldsymbol{x}}(k) - \boldsymbol{x})^\top P(k | \boldsymbol{x})$.
The Fisher information matrix (FIM),\cite{Fisher_1925} a central concept in parameter estimation, { quantifies the amount of information that the measurement outcomes provide about the parameters.} Its elements are defined as
$\mathbf{F}_{ij} = \sum_{k} {\partial_{x_i} P(k | \boldsymbol{x}) \, \partial_{x_j} P(k | \boldsymbol{x})}/{P(k | \boldsymbol{x})}$.
The precision of any unbiased estimator is limited by the Cramér-Rao bound, $\boldsymbol{V}_{\boldsymbol{x}} \geq (\nu  \mathbf{F} )^{-1}$, where $ \nu $ denotes the number of experimental runs. Since the FIM is derived from the probability distributions associated with a specific POVM, this bound is inherently dependent on the measurement scheme.
To establish the fundamental precision limit, independent of any particular measurement, the quantum Cramér-Rao bound (QCRB) based on the quantum Fisher information matrix (QFIM) is widely utilized.\cite{Helstrom1969,doi:10.1142/S0219749909004839}
The elements of the QFIM are defined as
$\mathbf{Q}_{ij} = \operatorname{Tr}[ \hat{\rho}_{\boldsymbol{x}} (\hat{L}_i \hat{L}_j + \hat{L}_j \hat{L}_i)/2 ]$, where the symmetric logarithmic derivative (SLD) $\hat{L}_i$ is defined via
$\partial_{x_i} \hat{\rho}_{\boldsymbol{x}} =  (\hat{\rho}_{\boldsymbol{x}} \hat{L}_i + \hat{L}_i \hat{\rho}_{\boldsymbol{x}})/2$. 
Thus, the QCRB can be expressed by the inequality
\begin{equation}
\boldsymbol{V}_{\boldsymbol{x}} \geq (\nu \mathbf{F})^{-1} \geq (\nu \mathbf{Q})^{-1}.
\end{equation}
Notably, equality in the second inequality is attainable if and only if the mean Uhlmann curvature matrix $\mathcal{U}$ vanishes. Its elements are given by
$\mathcal{U}_{ij} = i \operatorname{Tr}(\hat{\rho}_{\boldsymbol{x}}[\hat{L}_i,\hat{L}_j])/4.$ This condition, namely the weak commutativity condition,\cite{10.1063/1.2988130,PhysRevA.94.052108,Vidrighin2014,Carollo2018} is generally challenging to satisfy in multi-parameter estimation scenarios.

\begin{figure}[!t]
    \centering
    \includegraphics[width=0.7\linewidth]{diffusion_schematic6.pdf}
    \caption{Schematic of simultaneous estimation for phase $\phi$ and phase diffusion amplitude $\Delta$ by collective measurements. A random phase shift $\tilde{\phi}\sim\mathcal{N}(\phi, 2\Delta^2)$ is applied to one arm of the interferometer to encode both the parameter $\phi$ and $\Delta$. The final quantum state, $\hat{\rho}_{\phi,\Delta}$, is generated, and collective measurements are performed on two copies of this state, $\hat{\rho}_{\phi,\Delta}^{\otimes 2}$, to jointly estimate the parameters.}
    \label{schematic}
\end{figure}

It is important to emphasize that the SLD operators corresponding to the phase parameter $\phi$ and the phase diffusion amplitude $\Delta$ generally do not satisfy the weak commutativity condition. For equatorial states (i.e., when $\theta = \pi/2$), one obtains
$\operatorname{Tr}(\hat{\rho}_{\phi,\Delta}\,[\hat{L}_{\phi},\hat{L}_{\Delta}]) = 0,$
indicating that the QCRB for these parameters can, in principle, be asymptotically achieved. In this case, the QFIM is 
\begin{equation}
    \mathbf{Q} =\begin{bmatrix}
        e^{-2\Delta^2} & 0\\
         0 & \frac{4\Delta^2}{e^{2\Delta^2}-1}
    \end{bmatrix},
\end{equation}
and the derivation is provided in Appendix A.
In the case of mixed states, achieving the QCRB typically necessitates the implementation of collective measurements across multiple copies, as opposed to relying solely on measurements performed on individual copies.\cite{ PhysRevA.94.052108,10.1063/1.2988130,PhysRevLett.123.200503} Consequently, separable measurements are unable to saturate the QCRB, and an increased number of copies undergoing collective measurements results in enhanced precision of parameter estimation.\cite{Vidrighin2014,Roccia_2018}

To quantitatively assess the precision trade-off in the joint estimation of $\phi$ and $\Delta$,  we employ a figure of merit defined as 
$\operatorname{Tr}(\mathbf{Q}^{-1}\mathbf{F})$. Since the QFIM is diagonal, this figure of merit can be written in the form ${\mathbf{F}_{\phi\phi}}/{\mathbf{Q}_{\phi\phi}}+{\mathbf{F}_{\Delta\Delta}}/{\mathbf{Q}_{\Delta\Delta}}$. In the single-copy scenario, this figure of merit satisfies
$\operatorname{Tr}\left(\mathbf{Q}^{-1}\mathbf{F}\right) \leq 1.$ By extending this analysis to two-copy states, $\hat{\rho}_{\phi,\Delta}^{\otimes 2}$, we arrive at the following precision trade-off relation,
\begin{equation}\label{1.5bound}
\operatorname{Tr}\left(\mathbf{Q}_2^{-1}\mathbf{F}_2\right) \leq 1.5\,
\end{equation}
where $\mathbf{Q}_2=2\mathbf{Q}$ denotes the QFIM of $\hat{\rho}_{\phi,\Delta}^{\otimes 2}$, and $\mathbf{F}_2$ represents the FIM derived from the POVM applied to the two-copy quantum state. The details can be found in Appendix A and B. Furthermore, we demonstrate that employing  Bell measurements enable the figure of merit to saturate this bound in the limit as $\Delta \to 0$. Specifically, by projecting the two‐copy state $\hat{\rho}_{\phi,\Delta}^{\otimes 2}$ onto the four Bell bases $(|00\rangle + |11\rangle)/\sqrt{2},(|00\rangle - |11\rangle)/\sqrt{2},(|01\rangle + |10\rangle)/\sqrt{2},(|01\rangle + |10\rangle)/\sqrt{2}$
, the corresponding probabilities at the four output ports are given by,

\begin{equation}
\begin{aligned}\label{diffu_4output1}
p_1 &= \frac{1}{4} \left( 1 + e^{-2\Delta^2} \cos2\phi \right), &
p_3 &= \frac{1}{4} \left( 1 + e^{-2\Delta^2} \right), \\
p_2 &= \frac{1}{4} \left( 1 - e^{-2\Delta^2} \cos2\phi \right), &
p_4 &= \frac{1}{4} \left( 1 - e^{-2\Delta^2} \right).
\end{aligned}
\end{equation}
Therefore, for $\hat{\rho}_{\phi,\Delta}^{\otimes 2}$ under Bell measurements, the figure of merit becomes,
\begin{equation}\label{ratio}
\frac{\mathbf{F}_{2,\phi\phi}}{\mathbf{Q}_{2,\phi \phi}} + \frac{\mathbf{F}_{2,\Delta \Delta}}{\mathbf{Q}_{2,\Delta \Delta}} = \frac{1}{1+e^{2\Delta^2}} + \frac{1-2e^{2\Delta^2}+\cos4\phi}{1-2e^{4\Delta^2}+\cos4\phi}.    
\end{equation}
It follows that, in the limit $\Delta \to 0$, the precision bound given by Equation~\eqref{1.5bound} is saturated.

\section{Experiment setup and results}
\begin{figure}[!t]
    \centering
    \includegraphics[width=0.8\linewidth]{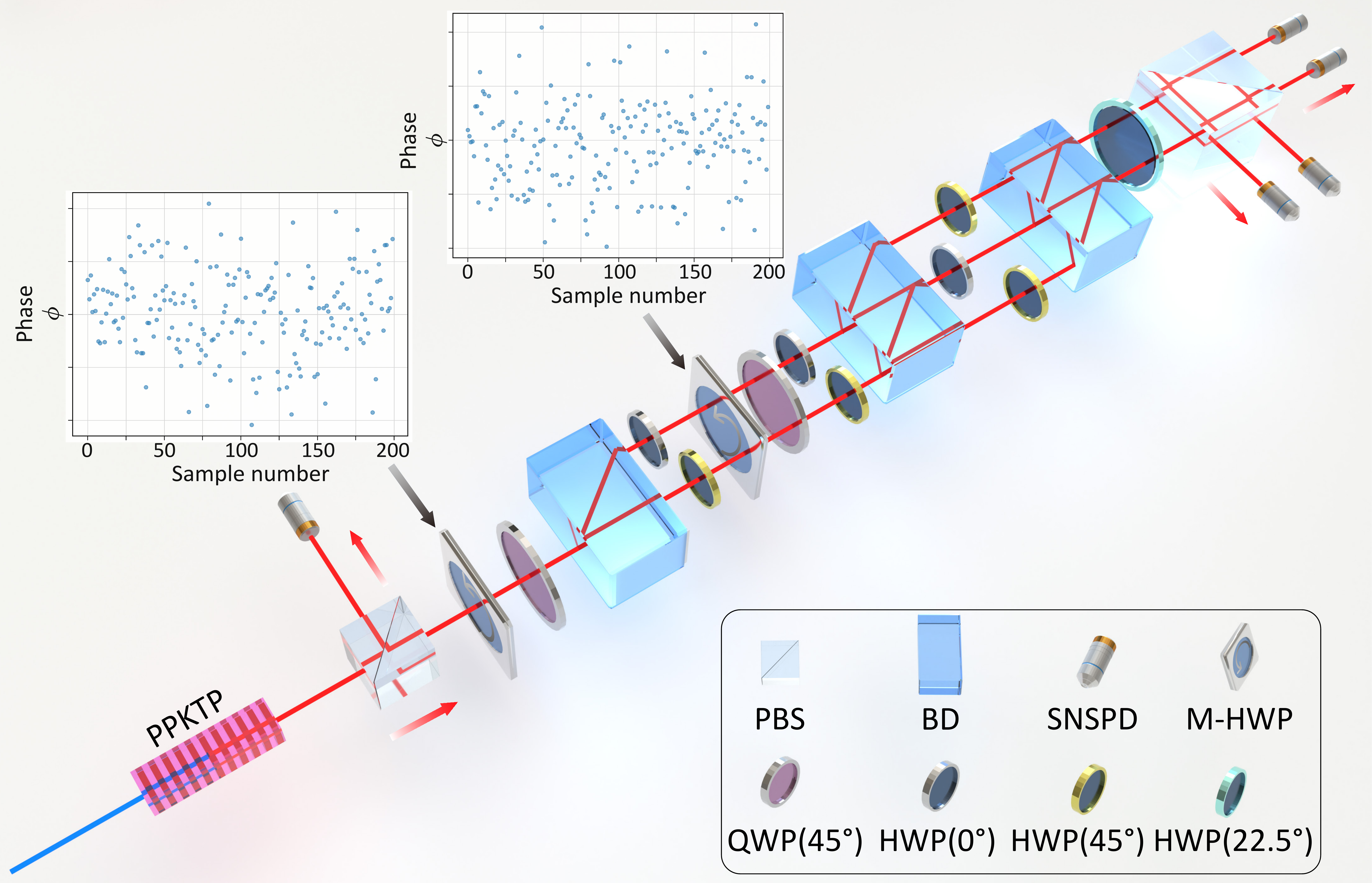}
    \caption{Experimental setup for the joint estimation of phase $\phi$ and phase diffusion $\Delta$ via Bell measurements. The setup utilizes a four-step quantum walk to prepare a two-copy parameterized quantum state and perform deterministic Bell measurements.}
    \label{fig:setup}
\end{figure}

The experimental setup depicted in \textbf{Figure}~\ref{fig:setup} is utilized to jointly estimate the parameters  $\phi$ and $\Delta$ via  deterministic Bell measurements. This setup involves preparing the parameterized two-copy state $\hat{\rho}_{\phi,\Delta}^{\otimes 2}$, followed by the deterministic Bell measurements.

A pair of 1560 nm photons is produced through type-II spontaneous parametric down-conversion in a periodically poled potassium titanyl phosphate (PPKTP) crystal. These photons are separated into distinct paths by a polarizing beam splitter (PBS). The vertically polarized (V) photon is detected by a superconducting nanowire single-photon detector (SNSPD) and serves as a heralding signal. Meanwhile, the horizontally polarized (H) photon is used to generate the parameterized two-copy state with parameters $\phi$ and $\Delta$.

In our approach, a four-step quantum walk is employed to implement both parameter encoding and Bell measurements. The first qubit is encoded in the photon's path degree of freedom (DOF), with the logical states  defined as $| \mathrm{up} \rangle \equiv | 0 \rangle$ and $| \mathrm{down} \rangle \equiv | 1 \rangle$. The second qubit is encoded in the polarization DOF with $| H \rangle \equiv | 0 \rangle$ and $| V \rangle \equiv | 1 \rangle$. To encode the parameters $\phi$ and $\Delta$ into the polarization DOF, the horizontally polarized photon $|H\rangle$ is first passed through a combination of a motorized half-wave plate (M-HWP) and a quarter-wave plate (QWP) fixed at a rotation angle of $\pi/4$. The M-HWP applies a sequence of discrete phase shifts $(\phi_1, \phi_2, \dots, \phi_{200})$ drawn from a Gaussian distribution $\mathcal{N}(\phi, 2\Delta^2)$, thereby encoding both $\phi$ and $\Delta$ into the polarization DOF. Subsequently, a beam displacer (BD) deflects the horizontal polarization $|H\rangle$ into the up path $| \mathrm{up} \rangle$, creating a 4 mm spatial separation from the vertical polarization $|V\rangle$ in the down path $|\mathrm{down}\rangle$. This operation effectively maps the polarization qubit onto a path qubit. Thereafter, half-wave plates (HWPs), set to $0$ and $\pi/4$ in the $| \mathrm{up} \rangle$ and $| \mathrm{down} \rangle$ paths respectively, are inserted to prepare the photon in the state $\hat{\rho}_{\phi,\Delta} \otimes | H \rangle \langle H |$. This state is subsequently processed through an M-HWP and a QWP, analogous to those used for encoding the first qubit, to encode the second polarization qubit. The complete procedure ultimately yields the parameterized two-copy quantum state $ \hat{\rho}_{\phi,\Delta}^{\otimes 2} $. A deterministic Bell measurement is then performed on this two-copy quantum state by constructing three specific coin operators and executing a three-step quantum walk (Appendix D). Finally, photon counts are recorded at four output ports using SNSPDs. Each output port corresponds to a projection onto one of the four Bell states, thereby enabling deterministic Bell state measurements.

\begin{figure}[t!]
    \centering
    \includegraphics[width=0.8\linewidth]{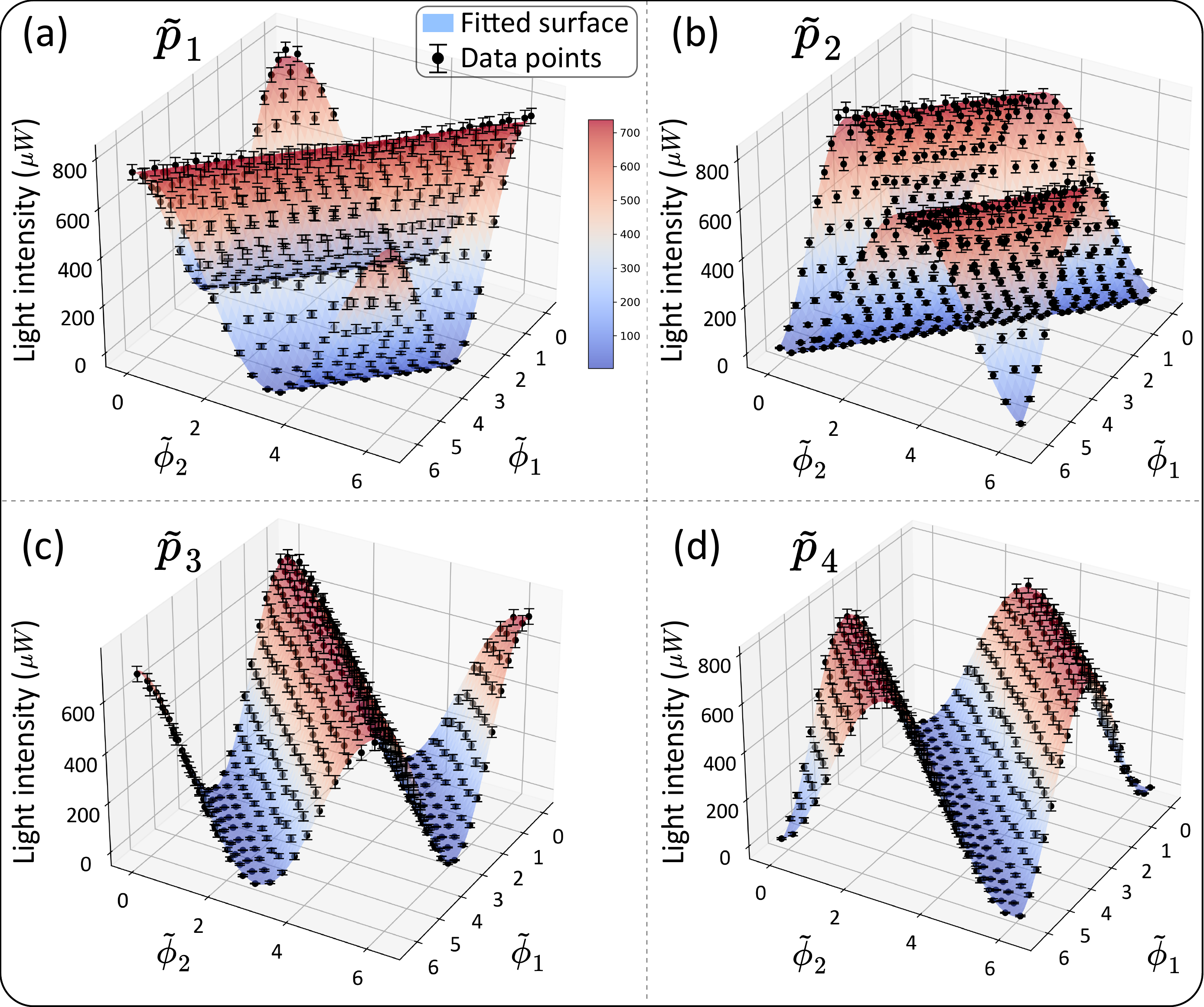}
    \caption{Distribution of the intensities at four output ports as functions of $\tilde{\phi}_1,\tilde{\phi}_2 \in [0,2\pi)$. Different colors indicate varying intensity values. Each surface in the plots is generated by fitting over a grid of $100\times100$ sampled data points within this interval.}
    \label{fig:likehood}
\end{figure}

In our experimental setup, both the phase and phase diffusion amplitude are simultaneously encoded by applying a large number of discrete phase values via the M-HWPs. This method requires stringent control over the phase implementation process. To ensure the precision of the apparatus, we perform a calibration and validation procedure using 1560 nm classical light. In this calibration, we use 100 uniformly discretized phase points spanning the entire interval $[0, 2\pi]$ to prepare the quantum state $\hat{\rho}(\tilde{\phi}_1) \otimes \hat{\rho}(\tilde{\phi}_2)$. A set of Bell measurements is then carried out on this state, with the light intensities detected at the four output ports. The corresponding probability distributions are
\begin{equation}
\begin{aligned}
\tilde{p}_1 &= \frac{1}{4}\Bigl(1 + \cos(\tilde{\phi}_1 + \tilde{\phi}_2)\Bigr), & \tilde{p}_2 &= \frac{1}{4}\Bigl(1 - \cos(\tilde{\phi}_1 + \tilde{\phi}_2)\Bigr), \\
\tilde{p}_3 &= \frac{1}{4}\Bigl(1 + \cos(\tilde{\phi}_1 - \tilde{\phi}_2)\Bigr), & \tilde{p}_4 &= \frac{1}{4}\Bigl(1 - \cos(\tilde{\phi}_1 - \tilde{\phi}_2)\Bigr).
\end{aligned}\label{eq:diffusion_likehood}
\end{equation}
For each output port, we record 10,000 intensity data points while varying $\tilde{\phi}_1$ and $\tilde{\phi}_2$ sequentially. The fitted surface curves representing the intensity distributions (as shown in \textbf{Figure}~\ref{fig:likehood}) agree well with the theoretical probability distributions in Equation~\eqref{eq:diffusion_likehood}, and the interference visibility at all four ports exceeds 99.7\%, thereby verifying the feasibility of the experimental apparatus.

\begin{figure}[t]
    \centering
    \includegraphics[width=0.8\linewidth]{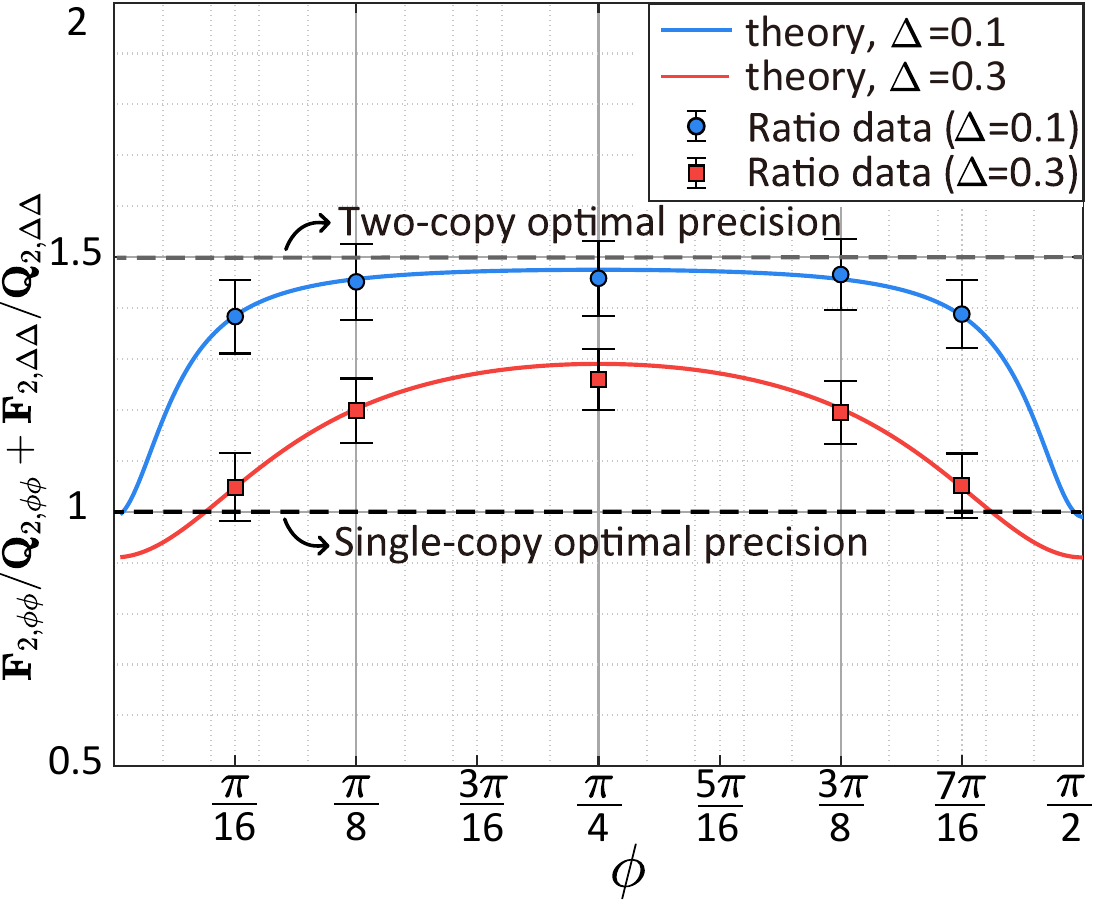}
    \caption{The experimental results of joint estimation of phase $\phi$ and phase diffusion amplitude $\Delta$ via Bell measurements. The joint estimation precision is evaluated across ten data sets with $\Delta \in \{0.1, 0.3\}$, $\phi \in \{\pi/16, \pi/8, \pi/4, 3\pi/8, 7\pi/16\}$ and $\nu \approx 10^{4}$.}
    \label{fig:diffusion_ratiovar}
\end{figure}

Thereafter, a 1560 nm single-photon source is coupled into the calibrated experimental setup. Two independent sets of discrete phases, each drawn from the same Gaussian distribution, are simultaneously applied to the two M-HWPs. In the experiment, 200 discrete phase points are implemented to encode the parameters and thereby prepare the two-copy quantum state 
$\hat{\rho}_{\phi,\Delta}^{\otimes 2}$. Photon counts are recorded at the four output ports using SNSPDs, yielding total photon counts $N_1$, $N_2$, $N_3$, and $N_4$ during the entire phase-loading process, with 
$\nu = N_1 + N_2 + N_3 + N_4 \approx 10^{4}$. Based on the four probability distributions given in Equation~(\ref{diffu_4output1}), we employ maximum likelihood estimation to obtain the estimators for the parameters $\phi$ and $\Delta$. The likelihood function is defined as $L = p_1^{N_1} \cdot p_2^{N_2} \cdot p_3^{N_3} \cdot p_4^{N_4}$. 
The values of $\phi$ and $\Delta$ that maximize $L$ are taken as the estimates, yielding one set of joint estimates for phase and phase diffusion. By repeating the above experimental procedure, we acquire 400 sets of joint estimates, from which the estimation variances and the covariance of the two parameters are determined to explore the ultimate precision limits achievable by this scheme.

\begin{figure}[t]
    \centering
    \includegraphics[width=\linewidth]{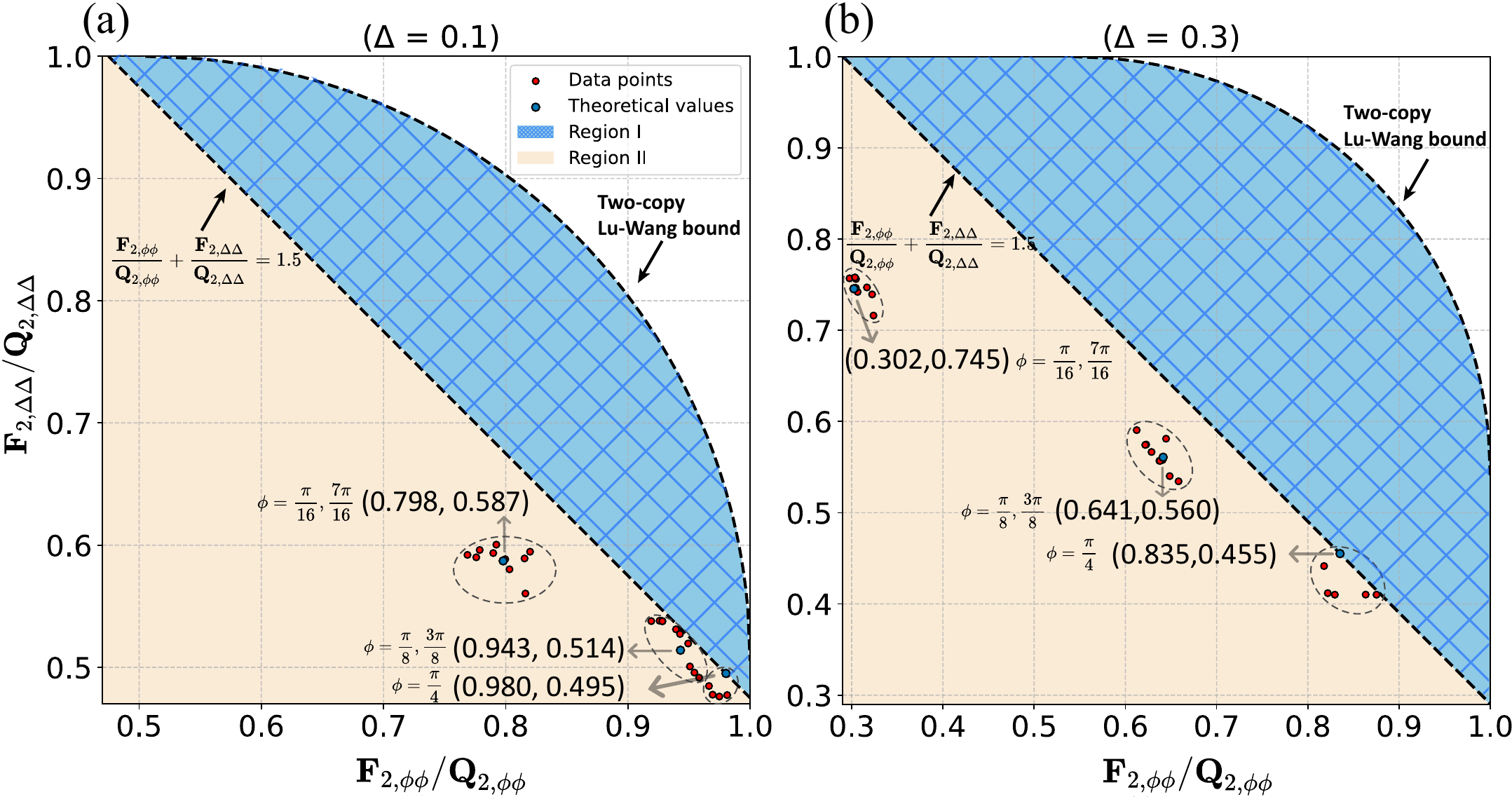}
    \caption{Comparison of the theoretical two-copy Lu--Wang bound in Eq.~(\ref{eq:Ratio_2copy}) with achievable precision limits in Eq.~(\ref{1.5bound}) for a two-qubit system. (a) $\Delta=0.1$. (b) $\Delta=0.3$. The plots depict the trade-off between phase and diffusion precision. Although the Lu--Wang uncertainty relation allows for both Regions I and II, only Region II is practically attainable. The 25 experimental data points in each plot confirm this, all lying within Region II.}
    \label{fig:LW_data}
\end{figure}

Since the precision of the joint estimation of phase and phase diffusion amplitude via Bell measurements depend on the parameters themselves, as shown in Equation~(\ref{ratio}), we experimentally demonstrate the joint estimation results for $\Delta \in \{0.1, 0.3\}$ and $\phi \in \{\pi/16,\,\pi/8,\,\pi/4,\,3\pi/8,\,7\pi/16\}$. These results are presented in \textbf{Figure}~\ref{fig:diffusion_ratiovar}. In this figure, the curves correspond to the theoretical prediction of the figure of merit $\operatorname{Tr}(\mathbf{F}_2 \mathbf{Q}_2^{-1})$ as a function of $\phi$. We estimate the Fisher information matrix from the covariance matrix acquired from the experiment results. The error bars on the experimental data points are determined via Monte Carlo simulations using 100 independent samples drawn from the corresponding Poisson distributions.\cite{Carolan2014,Vidrighin2014}
For $\Delta = 0.1$, the precision reaches its optimum at $\phi = \pi/4$, where the figure of merit saturates at 1.475. This verifies that the joint estimation of phase and phase diffusion via Bell measurements can achieve the optimal precision for two-copy quantum states. As the five blue data points show little variation, we further measure the variation of the figure of merit with $\phi$ for $\Delta = 0.3$. The experimental data agrees well with theoretical predictions, thereby validating our theoretical framework.

Lu and Wang extended Heisenberg's uncertainty principle to quantum multi-parameter estimation, arriving at a metrological bound on the simultaneous estimation of two parameters.\cite{PhysRevLett.126.120503} This result, known as the Lu--Wang uncertainty relation, establishes a strict precision limit for the joint estimation of phase and phase diffusion amplitude when using a single-copy state $\hat{\rho}_{\phi,\Delta}$.
However, when this bound is extended to the case of two-copy states, $\hat{\rho}_{\phi,\Delta}^{\otimes 2}$, we obtain the generalized inequality
\begin{equation}\label{eq:Ratio_2copy}
\begin{split}
    \frac{\mathbf{F}_{2,\phi\phi}}{\mathbf{Q}_{2,\phi \phi}} +
    \frac{\mathbf{F}_{2,\Delta \Delta}}{\mathbf{Q}_{2,\Delta \Delta}}
    - \sqrt{3-e^{-2\Delta^2}}
      \sqrt{1-\frac{\mathbf{F}_{2,\phi\phi}}{\mathbf{Q}_{2,\phi \phi}}}
      \sqrt{1-\frac{\mathbf{F}_{2,\Delta \Delta}}{\mathbf{Q}_{2,\Delta \Delta}}}
    \leq \frac{7 - e^{-2\Delta^2}}{4}\,.
\end{split}
\end{equation}
Notably, in this context the bound is not tight.
The precision trade-off between phase and phase diffusion, as imposed by the Lu--Wang uncertainty relation for $\Delta=0.1$ and $\Delta=0.3$, is illustrated in \textbf{Figure}~\ref{fig:LW_data}. The plots display the ratio 
$\mathbf{F}_{2,\phi \phi}/\mathbf{Q}_{2,\phi \phi}$ versus $\mathbf{F}_{2,\Delta \Delta}/\mathbf{Q}_{2,\Delta \Delta}$ according to the Equation~\eqref{ratio}. In each subplot, the blue hatched area (Region I) and the yellow area (Region II) together represent the theoretically allowed region dictated by the Lu--Wang uncertainty relation. The upper bound of Region I is given by Equation.~(\ref{eq:Ratio_2copy}). However, the practically achievable precision is confined to Region II, in which the boundary is limited by Equation.~(\ref{1.5bound}), and the range indicated by Region I remains unattainable in practice.
In the figure, blue dots represent the theoretical predictions for 
$\phi \in \left\{ \frac{\pi}{16},\allowbreak\, \frac{\pi}{8},\allowbreak\, \frac{\pi}{4},\allowbreak\, \frac{3\pi}{8},\allowbreak\, \frac{7\pi}{8} \right\}$ (which appear as three distinct points due to degenerate values), while red dots indicate the experimental data, with each subfigure containing 25 data points. 
The fact that all experimental data points fall within Region II demonstrates that the experimental outcomes are consistent with the theoretical predictions. This analysis confirms that, although the Lu--Wang uncertainty relation provides a corrected precision limit for mixed states, its application to the joint estimation of phase and phase diffusion in two-copy quantum states ultimately yields a bound that is not tight in practice.

\section{Conclusion and Discussions}

In summary, we have experimentally demonstrated that deterministic Bell measurements on a two-copy quantum system enable joint estimation of phase and phase diffusion with enhanced precision. By implementing quantum walks in a linear-optical network, we encode both parameters into a two-copy state and perform deterministic Bell measurements, achieving an approximate 50 \% improvement over separable measurement strategies. These results validate our theoretical framework and establish deterministic Bell measurements as a powerful tool for multi-parameter estimation. Our work provides a robust foundation for advancing precision measurement techniques in noisy quantum systems. 

The fundamental precision limit of parameter estimation is governed by the Holevo-Cramér-Rao bound, which typically requires collective measurements on multiple identical copies of quantum states.~\cite{demkowicz2020multi,szczykulska2016multi,liu2020quantum,albarelli2020perspective} In the phase and phase diffusion estimation problem, the Holevo-Cramér-Rao bound coincides with the quantum Cramér-Rao bound due to the satisfaction of the weak commutativity condition. In principle, the figure of merit used can reach a value of 2, which requires collective measurements on more copies of quantum states. Therefore, exploiting collective measurements across more copies can further enhance precision in phase and phase diffusion estimation and deserves further investigation.


\appendix

\titleformat{\section}[block] {\normalfont\Large\bfseries\raggedright}
  {Appendix~\thesection:}
  {1em}
  {}           

\section{Derivation of the Lu--Wang uncertainty relation for Phase and Phase-Diffusion Estimation in Single- and Two-Copy Qubit Systems}
In quantum multi-parameter estimation, the constraints imposed by the Heisenberg uncertainty principle prevent the simultaneous realization of optimal measurements for different parameters.\cite{BUSCH2007155,Heisenberg1927} Although the quantum Cramér–Rao bound (QCRB) defines the ultimate precision limit, it is not always attainable in practice. While the Holevo-Cramér–Rao bound, in principle, provides the best achievable lower bound, its evaluation demands a complex optimization over certain operators. To circumvent these challenges, Lu and Wang proposed a scheme that explicitly incorporates measurement uncertainty into quantum multi-parameter estimation by leveraging the Heisenberg uncertainty principle. Their approach yields a more obvious trade-off between the precisions of different parameters by defining the “information regret” (i.e., the unextracted quantum Fisher information) and linking it to the error–uncertainty relations established by Ozawa and Branciard,\cite{OZAWA2004367,PhysRevLett.110.220402,PhysRevA.90.042113} thereby leading to what is now known as the Lu--Wang uncertainty relation.

To quantitatively capture the unextracted quantum Fisher information in a quantum measurement, the information regret matrix for a given POVM $\{\hat{\Pi}\}$ is defined as
\begin{equation}\label{eq:regret}
    \mathbf{R}(\{\hat{\Pi}\}) = \mathbf{Q} - \mathbf{F}(\{\hat{\Pi}\}),
\end{equation}
where $\mathbf{Q}$ denotes the quantum Fisher information matrix (QFIM) and $\mathbf{F}(\{\hat{\Pi}\})$ represents the  Fisher information matrix (FIM) corresponding to the measurement $\{\hat{\Pi}\}$. Since both $\mathbf{Q}$ and $\mathbf{F}(\{\hat{\Pi}\})$ are real symmetric matrices, the regret matrix $\mathbf{R}(\{\hat{\Pi}\})$ is a real symmetric, positive semidefinite matrix.

In the single-parameter case, Braunstein and Caves have shown that there exists an optimal measurement satisfying $\mathbf{F}(\{\hat{\Pi}\}) = \mathbf{Q}$, which results in a vanishing regret matrix $\mathbf{R}(\{\hat{\Pi}\})$.\cite{PhysRevLett.72.3439} However, in the multi-parameter scenario, the optimal measurements for different parameters are generally incompatible, so the components of $\mathbf{R}(\{\hat{\Pi}\})$ cannot all be zero simultaneously. This is a manifestation of the Heisenberg uncertainty principle in multi-parameter estimation. A normalized square-root regret is then defined as
$\Delta_j = \sqrt{{\mathbf{R}_{jj}}/{\mathbf{Q}_{jj}}}$, with $\Delta_j$ taking values in the interval $[0,1]$. Based on this definition, Lu and Wang derived the precision trade-off relation (the Lu--Wang uncertainty relation) between two parameters $x_j$ and $x_k$:
\begin{equation}\label{eq:tradeoff_bound}
    \Delta_j^2 + \Delta_k^2 + 2\sqrt{1-c_{jk}^2}\,\Delta_j\,\Delta_k \geq c_{jk}^2,
\end{equation}
where the coefficient $c_{jk}$ is defined as
\begin{equation}\label{eq:cjk}
    c_{jk} = \frac{|\operatorname{Im} \,\boldsymbol{\Omega}_{jk}|}{\sqrt{\operatorname{Re}\,\boldsymbol{\Omega}_{jj}\operatorname{Re}\,\boldsymbol{\Omega}_{kk}}}
    = \frac{|\operatorname{Im}\,\boldsymbol{\Omega}_{jk}|}{\sqrt{\mathbf{Q}_{jj}\mathbf{Q}_{kk}}},
\end{equation}
with $\boldsymbol{\Omega}_{jk} = \operatorname{Tr}(\hat{L}_j \hat{L}_k \hat{\rho}_{\boldsymbol{x}})$ and $\hat{L}_j$ being the symmetric logarithmic derivative (SLD) operator with respect to parameter $x_j$. For $c_{jk} \neq 0$, Equation~(\ref{eq:tradeoff_bound}) delineates the trade-off between the measurement regrets of the corresponding parameters. For pure states, the bound is tight and for mixed states, one can introduce the modified coefficient
\begin{equation}\label{eq:xiucjk}
     \tilde{c}_{jk} = \frac{\operatorname{Tr} \left| \sqrt{\hat{\rho}_{\boldsymbol{x}}} [\hat{L}_j, \hat{L}_k] \sqrt{\hat{\rho}_{\boldsymbol{x}}} \right|}{2\sqrt{\mathbf{Q}_{jj}\mathbf{Q}_{kk}}},
\end{equation}
where $|X| = \sqrt{X^\dagger X}$. For any quantum state, $\tilde{c}_{jk}$ is no less than $c_{jk}$, and for all pure states, $\tilde{c}_{jk} = c_{jk}$. An alternative form of the trade-off relation, based on the estimation error, is given by
\begin{equation}\label{eq:error_tradeoff}
    \gamma_j + \gamma_k - 2\sqrt{1-\tilde{c}_{jk}^2}\,\sqrt{(1-\gamma_j)(1-\gamma_k)} \leq 2 - \tilde{c}_{jk}^2,
\end{equation}
where, for brevity, we define $\gamma_j \equiv 1/(\nu \,\mathcal{E}_{jj}\,\mathbf{Q}_{jj})$, with $\mathcal{E}$ being the covariance matrix of the parameters and $\nu$ the number of experimental repetitions. This inequality is obtained by combining Equation~(\ref{eq:tradeoff_bound}) with the QCRB $\mathcal{E}_{jj} \geq  (\mathbf{F}^{-1})_{jj} /\nu \geq 1/(\nu\,\mathbf{Q}_{jj})$. These results reveal the error bounds that arise from measurement incompatibility in quantum multi-parameter estimation.

We now apply the Lu--Wang uncertainty relation to the specific problem of jointly estimating phase and phase diffusion. 
 Before that, we need to calculate the SLD operators and the QFIM of our model, which is also presented in the supplemental material of Ref.~\cite{PhysRevLett.126.120503}
First, the single-copy two-level quantum
system with phase and phase diffusion is written by
\begin{equation}
\hat{\rho}_{\phi,\Delta} = \frac{1}{2}\begin{pmatrix}
1 &  e^{-i\phi - \Delta^2} \\
 e^{i\phi - \Delta^2} & 1
\end{pmatrix}.
\end{equation}
After solving the equation $\partial_\kappa\hat{\rho}_{\phi,\Delta}=\hat{L}_\kappa\hat{\rho}_{\phi,\Delta}+\hat{\rho}_{\phi,\Delta}\hat{L}_\kappa$ with $\kappa=\phi$ or $\Delta$, the SLD operators corresponding to the parameters $\phi$ and $\Delta$ are found to be
\begin{align} \label{GMT}
    \hat{L}_{\phi} =
    \begin{pmatrix}
    0 & -i e^{-{\Delta}^2-i\phi} \\
    i e^{-{\Delta}^2+i\phi} & 0
    \end{pmatrix},   
    ~\hat{L}_{\Delta} = \frac{1}{e^{2{\Delta}^2}-1}\,
    \begin{pmatrix}
    2\Delta & -2\Delta\, e^{{\Delta}^2-i\phi} \\
    -2\Delta\, e^{{\Delta}^2+i\phi} & 2\Delta
    \end{pmatrix}.
\end{align}
Accordingly, the matrix $\boldsymbol{\Omega}$ can be written as
\begin{equation}\label{Omega}
    \boldsymbol{\Omega} = \begin{pmatrix}
    e^{-2{\Delta}^2} & 0 \\
    0 & \dfrac{4\Delta^2}{e^{2{\Delta}^2}-1}
    \end{pmatrix}.
\end{equation}
Since the imaginary part of $\Omega$ vanishes, the QFIM $\mathbf{Q}$ equals $\Omega$ and the weak commutativity condition is satisfied.
It is straightforward from Equation~(\ref{eq:cjk}) that $c_{12} = 0$. However, since the quantum state $\hat{\rho}$ in the phase and phase diffusion problem is mixed, to further characterize the incompatibility between the parameters one computes the modified coefficient $\widetilde{c}_{12}$ using Equation~(\ref{eq:xiucjk}). The numerator is given by
\begin{equation}
  \operatorname{Tr} \left| \sqrt{\hat{\rho}} \, [\hat{L}_{\phi}, \hat{L}_{\Delta}] \, \sqrt{\hat{\rho}} \right| = \frac{4 e^{-\Delta^2} \Delta}{\sqrt{e^{2\Delta^2} - 1}}.
\end{equation}
Taking into account the diagonal elements in Equation~(\ref{Omega}), one finally obtains
\begin{equation}
    \widetilde{c}_{12} = \frac{\operatorname{Tr} \left| \sqrt{\hat{\rho}} \, [\hat{L}_{\phi}, \hat{L}_{\Delta}] \, \sqrt{\hat{\rho}} \right|}{2\sqrt{\mathrm{Re}\, \boldsymbol{\Omega}_{11}\,\mathrm{Re}\, \boldsymbol{\Omega}_{22}}} = 1.
\end{equation}
Substituting $\widetilde{c}_{12} = 1$ into the Lu--Wang uncertainty relation in Equation~(\ref{eq:tradeoff_bound}), we obtain
\begin{equation}
    \frac{\mathbf{F}_{\phi \phi}}{\mathbf{Q}_{\phi \phi}} + \frac{\mathbf{F}_{\Delta \Delta}}{\mathbf{Q}_{\Delta \Delta}} \le 1.
\end{equation}


For the joint estimation of phase and phase diffusion in two-copy qubit systems, the Lu--Wang uncertainty relation has not yet been discussed. Here, we extend the bound to a two-qubit system in order to investigate the precision limits it establishes.
In this work we do not directly follow the approach which diagonalizes the two-copy quantum state $\hat{\rho}^{\otimes2}$ to obtain the SLD operators $\hat{L}^{\otimes2}_\phi$ and $\hat{L}^{\otimes2}_\Delta$—but instead use the relation
\begin{equation}
\begin{aligned}
\partial (\hat{\rho}^{\otimes2}) &= \partial (\hat{\rho} \otimes \hat{\rho}) = \partial \hat{\rho} \otimes \hat{\rho} + \hat{\rho} \otimes \partial \hat{\rho} \\
&= \frac{1}{2} \left[ (\hat{\rho} \otimes \hat{\rho})\cdot(\hat{L} \otimes \mathbf{I} + \mathbf{I} \otimes \hat{L})  + (\hat{L} \otimes \mathbf{I} + \mathbf{I} \otimes \hat{L})\cdot(\hat{\rho} \otimes \hat{\rho}) \right] \\
&= \frac{1}{2} \left[ (\hat{\rho}^{\otimes2})\cdot\mathbf{\hat{L}} + \hat{\mathbf{L}}\cdot(\hat{\rho}^{\otimes2}) \right],
\end{aligned}
\end{equation}
where $\hat{\rho}$, $\hat{L}$, and $\mathbf{I}$ correspond to the operators in the qubit system. By the definition of the SLD operator, the SLD for the two-copy state $\hat{\rho}^{\otimes2}$ is given by
\begin{equation}
\hat{\mathbf{L}} \equiv \hat{L} \otimes \mathbf{I} + \mathbf{I} \otimes \hat{L}.
\end{equation}
Thus, the explicit forms of $\hat{L}^{\otimes2}_\phi$ and $\hat{L}^{\otimes2}_\Delta$ can be computed as
\begin{equation}
\begin{aligned}
\hat{\mathbf{L}}_\phi &= \begin{pmatrix}
0 & -i\,e^{-\Delta^2-i\phi} & -i\,e^{-\Delta^2-i\phi} & 0 \\
i\,e^{-\Delta^2+i\phi} & 0 & 0 & -i\,e^{-\Delta^2-i\phi} \\
i\,e^{-\Delta^2+i\phi} & 0 & 0 & -i\,e^{-\Delta^2-i\phi} \\
0 & i\,e^{-\Delta^2+i\phi} & i\,e^{-\Delta^2+i\phi} & 0
\end{pmatrix}, \\ 
\hat{\mathbf{L}}_\Delta &= \frac{2\Delta}{e^{2\Delta^2}-1}
\begin{pmatrix}
2 & -e^{\Delta^2-i\phi} & -e^{\Delta^2-i\phi} & 0 \\
-e^{\Delta^2+i\phi} & 2 & 0 & -e^{\Delta^2-i\phi} \\
-e^{\Delta^2+i\phi} & 0 & 2 & -e^{\Delta^2-i\phi} \\
0 & -e^{\Delta^2+i\phi} & -e^{\Delta^2+i\phi} & 2
\end{pmatrix}.
\end{aligned}
\end{equation}
Combining these results with Equation~(\ref{GMT}), one can compute the modified coefficient for the mixed state 
$\widetilde{c}^{\otimes2}_{12} = {\sqrt{1+e^{-2\Delta^2}}}/{2}.$ 
Substituting this result into Equation~(\ref{eq:error_tradeoff}), the Lu--Wang uncertainty relation for the two-copy quantum state $\hat{\rho}^{\otimes2}(\phi,\Delta)$ takes the explicit form
\begin{equation}
\begin{split}
    \frac{\mathbf{F}_{2,\phi\phi}}{\mathbf{Q}_{2,\phi \phi}} + \frac{\mathbf{F}_{2,\Delta \Delta}}{\mathbf{Q}_{2,\Delta \Delta}} - \sqrt{3-e^{-2\Delta^2}}\sqrt{1-\frac{\mathbf{F}_{2,\phi\phi}}{\mathbf{Q}_{2,\phi \phi}}} 
    \times \sqrt{1-\frac{\mathbf{F}_{2,\Delta \Delta}}{\mathbf{Q}_{2,\Delta \Delta}}} \leq \frac{7 - e^{-2\Delta^2}}{4}\,.
\end{split}
\end{equation}
In practice, for the joint estimation of phase and phase diffusion in two-copy qubit systems the precision satisfies
$\operatorname{Tr}(\mathbf{Q}_2^{-1}\mathbf{F}_2) \leq 1.5,$ and the Lu--Wang uncertainty relation in this case is not tight, meaning that there exists a region of unattainable precision.

\section{ Precision Limits for Joint Estimation of Phase and Phase Diffusion in Two-Copy Qubit Systems}

Zhu and Hayashi derived a fundamental constraint on the  FIM for collective measurements on two-copy mixed states.\cite{PhysRevLett.120.030404} This constraint is stated in the form of a theorem as follows. For any POVM $\{ \hat{\Pi} \}$ acting on the Hilbert space $\mathcal{H}^{\otimes 2}$, the FIM $\mathbf{F}_2$ obtained from the quantum state ${\hat{\rho}_{\boldsymbol{x}}}^{\otimes 2}$ satisfies the inequality
\begin{equation}
    \operatorname{Tr}\Bigl(\mathbf{Q}^{-1}\mathbf{F}_2\Bigr) \le 3d - 3,
    \label{eq:ZhuHayashiBound}
\end{equation}
where $\mathbf{Q}$ is the QFIM of the single-copy quantum state ${\hat{\rho}_{\boldsymbol{x}}}$, and $d$ denotes the dimension of the single-system Hilbert space $\mathcal{H}$.

This expression constitutes a fundamental limit on the amount of information that can be extracted via collective measurements on any two-copy quantum state. By contrast, for separable measurements the FIM satisfies
\begin{equation}   \operatorname{Tr}\Bigl(\mathbf{Q}^{-1}\mathbf{F}_{2,\text{sep}}\Bigr) \le 2(d-1).
\end{equation}
It can be seen that the theoretical upper bound for entangled measurements is 50\% higher than that for separable measurements. This result rigorously and quantitatively reveals, from an information-theoretic perspective, the intrinsic advantage of entangled measurements over separable ones in terms of information extraction capability.
By substituting 
$\mathbf{Q}^{-1} = (\mathbf{Q}_2/2)^{-1}$ into Equation~\eqref{eq:ZhuHayashiBound}, one obtains
\begin{equation}\label{2copyD}    \operatorname{Tr}\Bigl(\mathbf{Q}_2^{-1}  \mathbf{F}_2\Bigr) \leq 1.5.
\end{equation}
This result establishes the precision limit for quantum state estimation of qubit states, which consists of three parameters. In the following, we will show the tight bound for jointly estimating the phase and phase diffusion, which is coincident with Eq.~(\ref{2copyD}).

We employ the precision trade-off relation for two-parameter estimation, as discussed in Yuan's work\cite{Chen2024}, to prove Equation~(\ref{2copyD}). That work demonstrates that, if the QFIM is diagonal, the precision for a two-parameter estimation satisfies
\begin{align}
  w_1\,(1 - \frac{\mathbf{F}_{11}}{\mathbf{Q}_{11}})
  +\,w_2\,(1 - \frac{\mathbf{F}_{22}}{\mathbf{Q}_{22}})
  \ge \sum_{q}\frac{\lambda_q}{2}
      \bigl(\alpha_q - \sqrt{\alpha_q^2 - \beta_q^2}\bigr)\,,
  \label{eq:tighter_tradeoff_2param}
\end{align}
where $w_1$ and $w_2$ are weight factors, and the parameters $\lambda_q$, $\alpha_q$, and $\beta_q$ are defined as follows. Any mixed state $\hat{\rho}_{\boldsymbol{x}}$ can be decomposed as an ensemble of pure states:
\begin{equation}
    \hat{\rho}_{\boldsymbol{x}} = \sum_q \lambda_q\, |\phi_q\rangle\langle\phi_q|,
\end{equation}
with $\lambda_q = \langle u_q | \hat{\rho}_{\boldsymbol{x}} | u_q \rangle$ and $|\phi_q\rangle ={\sqrt{{\hat{\rho}_{\boldsymbol{x}}}/\langle u_q|\hat{\rho}_{\boldsymbol{x}}|u_q\rangle}}|u_q\rangle$, 
where $\{|u_q\rangle\}$ defines a complete orthonormal basis. Based on this decomposition, the parameters $\alpha_q$ and $\beta_q$ are given by
\begin{align}
    \alpha_q = w_1\left( \Delta_{|\phi_q\rangle}\hat{L}_1 \right)^2 + w_2\left( \Delta_{|\phi_q\rangle}\hat{L}_2 \right)^2, 
    ~\beta_q  = i\,\sqrt{w_1 w_2}\,\langle\phi_q| [\hat{L}_1, \hat{L}_2] |\phi_q\rangle,
    \label{eq:beta_q_def}
\end{align}
where $\Delta_{|\phi_q\rangle}\hat{L}_{1/2}$ denotes the standard deviation of the SLD operator $\hat{L}_{1/2}$ in the pure state $|\phi_q\rangle$. For the joint estimation of phase and phase diffusion in a two-copy system, we set $w_1 = 0.5,\ w_2 = 0.5$ and choose the basis states as
$|u_{1}\rangle = (0,1,0,0)^{\top},|u_{2}\rangle = (0,0,1,0)^{\top}, |u_{3}\rangle = (1,0,0,0)^{\top}, |u_{4}\rangle = (0,0,0,1)^{\top}.$
Substituting these into Equation~(\ref{eq:tighter_tradeoff_2param}) again yields $\operatorname{Tr}(\mathbf{Q}_2^{-1} \mathbf{F}_2) \le 1.5$.   

\section{Optimal Measurement Scheme Based on Bell Measurements}

In this section, we demonstrate that Bell measurements can saturate the precision limit given by Equation~(\ref{2copyD}). Since the phase $\phi$ and phase diffusion amplitude $\Delta$ associated with states lying on the equatorial plane of the Bloch sphere satisfy the weak commutativity condition, we consider the two-copy quantum state at $\theta = \pi/2$, for which the density matrix is given by  
\begin{equation}
\hat{\rho}_{\phi,\Delta}^{\otimes2} = \frac{1}{4}
\begin{pmatrix}
1 & e^{-\Delta^2-i\phi} & e^{-\Delta^2-i\phi} & e^{-2\Delta^2-2i\phi} \\
e^{-\Delta^2+i\phi} & 1 & e^{-2\Delta^2} & e^{-\Delta^2-i\phi} \\
e^{-\Delta^2+i\phi} & e^{-2\Delta^2} & 1 & e^{-\Delta^2-i\phi} \\
e^{-2\Delta^2+2i\phi} & e^{-\Delta^2+i\phi} & e^{-\Delta^2+i\phi} & 1
\end{pmatrix}\,.
\end{equation}

We implement a Bell measurement scheme on the quantum state. The corresponding Bell basis states are defined as follows:
\begin{equation}\label{Bell base}
\begin{aligned}
|\Psi_1\rangle &= |\Phi^+\rangle 
  = \tfrac1{\sqrt2}\bigl(|00\rangle + |11\rangle\bigr),
~|\Psi_2\rangle = |\Phi^-\rangle 
  = \tfrac1{\sqrt2}\bigl(|00\rangle - |11\rangle\bigr),\\
|\Psi_3\rangle &= |\Psi^+\rangle 
  = \tfrac1{\sqrt2}\bigl(|01\rangle + |10\rangle\bigr),
~|\Psi_4\rangle = |\Psi^-\rangle 
  = \tfrac1{\sqrt2}\bigl(|01\rangle - |10\rangle\bigr).
\end{aligned}
\end{equation}
Using the above Bell states as elements of the POVM, i.e., $\hat{\Pi}^B_i = |\Psi_i\rangle\langle \Psi_i|$, the probability for each outcome is
$p_i = \operatorname{Tr}(\hat{\rho}_{\phi,\Delta}^{\otimes2}\,\hat{\Pi}^B_i).$
Thus, the probability distribution over the four output ports is given by

\begin{equation}\label{diffu_4output}
\begin{aligned}
p_1 &= \tfrac14\bigl(1 + e^{-2\Delta^2}\cos2\phi\bigr), &
p_2 &= \tfrac14\bigl(1 - e^{-2\Delta^2}\cos2\phi\bigr),\\
p_3 &= \tfrac14\bigl(1 + e^{-2\Delta^2}\bigr),          &
p_4 &= \tfrac14\bigl(1 - e^{-2\Delta^2}\bigr).
\end{aligned}
\end{equation}

\begin{figure}[t]
    \centering
    \includegraphics[width=0.8\linewidth]{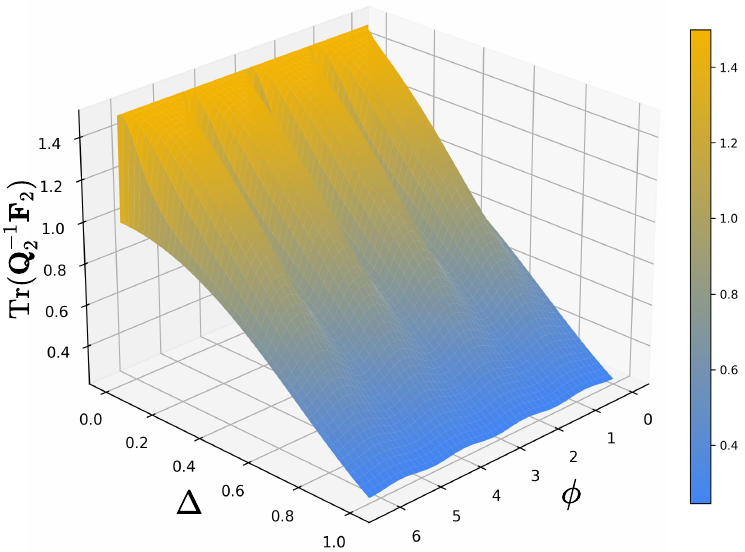}
  \caption{Three-dimensional visualization of the figure of merit $\operatorname{Tr}(\mathbf{Q}_2^{-1} \mathbf{F}_2)$ as a function of $\phi$ (ranging from 0 to $2\pi$) and $\Delta$ (ranging from 0 to 1).}
    \label{Ratio3}
\end{figure}

Based on these output probabilities, we further compute the FIM $\mathbf{F}_2$ for the two-copy state under the Bell measurements by, 
\begin{equation}
\mathbf{F}_2
\mkern-6mu=\mkern-6mu
\left(\!
  \begin{array}{@{}cc@{}}
    -\tfrac{4\sin^2(2\phi)}{1-2e^{4\Delta^2}+\cos4\phi}
    &
    -\tfrac{4\Delta\sin4\phi}{1-2e^{4\Delta^2}+\cos4\phi}
    \\[1ex]
    -\tfrac{4\Delta\sin4\phi}{1-2e^{4\Delta^2}+\cos4\phi}
    &
    \tfrac{4\Delta^2\,\bigl(-4\cos^2(2\phi)+e^{4\Delta^2}(3+\cos4\phi)\bigr)}
          {(e^{4\Delta^2}-1)\,(e^{4\Delta^2}-\cos^2(2\phi))}
  \end{array}
\!\right).
\end{equation}
Exploiting the property that the QFIM obeys
$\mathbf{Q}\left(\hat{\rho}\otimes\hat{\rho}\right) = 2\,\mathbf{Q}\left(\hat{\rho}\right),$
the QFIM for the two-copy state $\hat{\rho}_{\phi,\Delta}^{\otimes2}$ reads
\begin{equation}\label{QIF_33}
\mathbf{Q}_2 =
\begin{pmatrix}
2e^{-2\Delta^2} & 0 \\[2mm]
0 & \dfrac{8\Delta^2}{e^{2\Delta^2}-1}
\end{pmatrix}\,.
\end{equation}
Hence, the figure of merit, defined as
$\operatorname{Tr}(\mathbf{Q}_2^{-1}\mathbf{F}_2),$
can be expressed as
\begin{equation}
\frac{\mathbf{F}_{2,\phi\phi}}{\mathbf{Q}_{2,\phi \phi}} + \frac{\mathbf{F}_{2,\Delta \Delta}}{\mathbf{Q}_{2,\Delta \Delta}}
= \frac{1}{1 + e^{2\Delta^2}} + \frac{1 - 2e^{2\Delta^2} + \cos(4\phi)}{1 - 2e^{4\Delta^2} + \cos(4\phi)}\,.
\end{equation}
This equation reveals the explicit trade-off between the estimation precisions of $\phi$ and $\Delta$ when these parameters are simultaneously estimated via Bell measurements on the two-copy quantum state. A three-dimensional plot of the figure of merit as a function of $\phi$ and $\Delta$ is illustrated in \textbf{Figure}~\ref{Ratio3}. It is evident that as $\Delta$ approaches zero, the figure of merit tends toward 1.5. This indicates that the scheme achieves the optimal precision for the joint estimation of the two parameters using two-copy quantum states—a 50\% improvement over the separable measurement scheme based on single-copy states. 

\section{Deterministic Bell Measurements via Quantum Walk}

In a one-dimensional discrete quantum walk, the system state can be written as $ |x, c\rangle $, where $x = \dots,  -1,$ $ 0, 1, \dots$ denotes the walker’s position and $c = 0, 1$ denotes the coin state. The dynamics at each step are described by a unitary operator of the form
$U(t) = \hat{T}\, \hat{C}(t),$
where the conditional translation operator $\hat{T}$ is defined as
\begin{equation}
\hat{T} = \sum_{x}\Bigl(|x+1,0\rangle\langle x,0| + |x-1,1\rangle\langle x,1|\Bigr),
\end{equation}
and the coin operator $ \hat{C}(t) $ is given by
\begin{equation}
\hat{C}(t) = \sum_{x}|x\rangle\langle x| \otimes \hat{C}(x,t),
\end{equation}
with $ \hat{C}(x,t) $ being the coin operator associated with position $x$. By carefully designing the coin operators $ \hat{C}(x,t) $ at each position, one can implement a general POVM by measuring the walker's position after a predetermined number of steps.
 
\begin{figure}[t]
    \centering
    \includegraphics[width=0.8\linewidth]{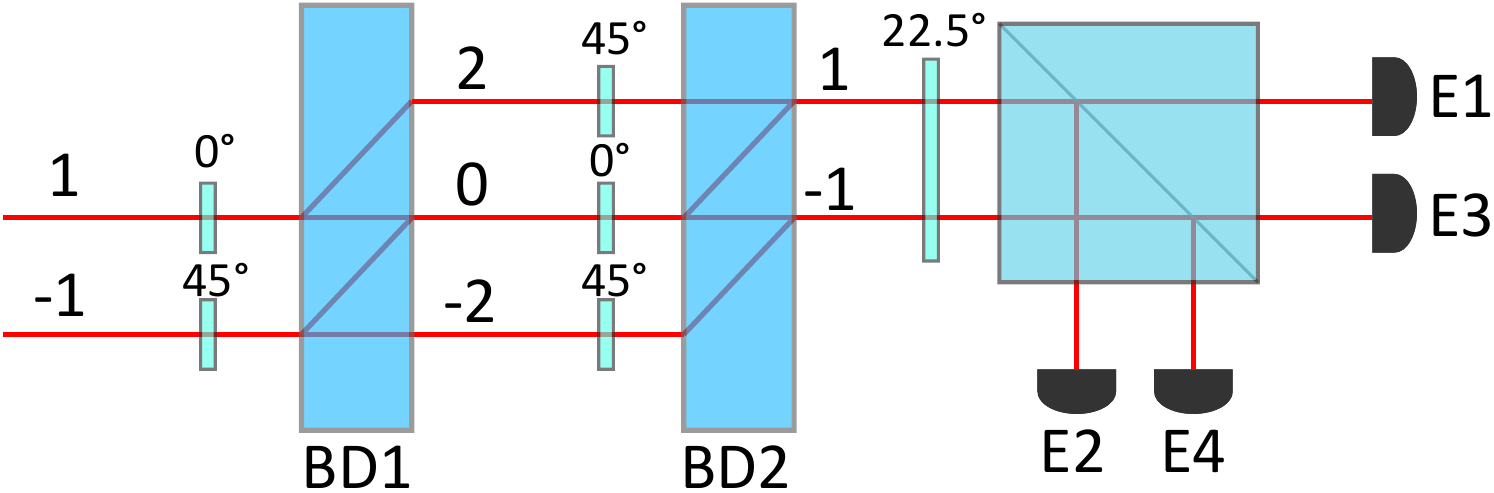}
  \caption{Schematic of the experimental setup for the deterministic Bell measurements using a three-step quantum walk.}
    \label{BellM}
\end{figure}

To realize a deterministic Bell measurement on the two-copy quantum state, we construct a scheme based on a three-step quantum walk, as shown in \textbf{Figure}~\ref{BellM}. In this scheme, the photon's path degree of freedom plays the role of the walker’s position while its polarization degree of freedom serves as the coin. Specifically, the upper path $|up\rangle$ is mapped to the walker position $x=1$ and the lower path $|down\rangle$ is mapped to $x=-1$. In addition, the polarization states $|H\rangle$ and $|V\rangle$ correspond to the coin states $c=0$ and $c=1$, respectively. Thus, the two-copy quantum state input into the Bell measurement apparatus can be regarded as the initial state of the quantum walk, distributed over the positions $x=\pm 1$. The entire evolution consists of three steps,  each comprising a coin operation followed by a conditional translation.

In the first step, different coin operations are applied to the photons located at $x=1$ and $x=-1$. These operations are implemented by half-wave plates (HWPs) and correspond to the coin operators
\begin{equation}
\hat{C}(1,1) = \begin{pmatrix} 1 & 0 \\ 0 & -1 \end{pmatrix}, \quad
\hat{C}(-1,1) = \begin{pmatrix} 0 & 1 \\ 1 & 0 \end{pmatrix}.
\end{equation}
This applies a Pauli-Z operator at $x=1$ and a Pauli-X operator at $x=-1$. Subsequently, the photon passes through the first beam displacer (BD1), which enacts the conditional translation $\hat{T}$. This shifts the coin state $|H\rangle$ ($c=0$) to position $x+1$ and the coin state $|V\rangle$ ($c=1$) to position $x-1$. After this step, the photon’s spatial position spreads to $x \in \{0,\pm2\}$.

In the second step, based on the new positions of the photon (now possibly at $x=0, 2,$ or $-2$), a position-dependent coin operation is applied using specific HWPs and quarter-wave plates (QWPs), realizing the coin operators
\begin{equation}
\hat{C}(0,2) = \begin{pmatrix} 1 & 0 \\ 0 & -1 \end{pmatrix}, \quad
\hat{C}(2,2) = C(-2,2) = \begin{pmatrix} 0 & 1 \\ 1 & 0 \end{pmatrix}.
\end{equation}
This means a Pauli-Z operator is applied at the central position $x=0$, while Pauli-X operators are implemented at the edge positions $x=\pm2$. Following this, the photon passes through a second beam displacer (BD2), which performs the conditional translation.

In the third and final step, the photon’s position evolves to $x \in \{\pm 1, \pm 3\}$. At this stage, photons originating from different initial paths can occupy the same spatial locations at $x = \pm 1$, resulting in spatial mode overlap. Taking advantage of this overlap, coin operations are applied only to photons at the central positions $x=\pm 1$, using the Hadamard operator
\begin{equation}
\hat{C}(1,3) = \hat{C}(-1,3) = \frac{1}{\sqrt{2}}\begin{pmatrix} 1 & 1 \\ 1 & -1 \end{pmatrix},
\end{equation}
which is implemented by an HWP rotated by $22.5^\circ$. This spatial overlap is crucial as it enables the interference necessary for the subsequent measurement to project onto the Bell states. No further translation is applied, and instead, a final projective measurement is performed using a polarizing beam splitter and four superconducting nanowire single-photon detectors to collect photon counts from four output ports. These measurement outcomes correspond to projections onto the four Bell basis states $\{ |\Psi^{+}\rangle,\, |\Psi^{-}\rangle,\, $
$|\Phi^{+}\rangle,\, |\Phi^{-}\rangle \}$ of the initial two-copy state. By recording the photon counts at the different detectors, the probability distribution of the Bell measurement can be reconstructed, thereby completing the joint estimation of the parameters $\phi$ and $\Delta$.

\section{Numerical Optimal Measurements Search for Large Diffusion}
As we proved in the main text, for small diffusion, the Bell measurements are nearly optimal in joint phase and phase estimation. However, for large diffusion, it is not optimal. Previous works~\cite{Vidrighin2014,Roccia_2018} numerically searched for the projective measurements on two-copy states, while the projective measurements are generally not ensured to be optimal.~\cite{matsumoto2002new}

The method for identifying the optimal positive operator-valued measure (POVM) is based on the gradient descent algorithm proposed by Zhang et al. ~\cite{zhang2024qestoptpovmiterativealgorithmoptimal}. As a representative example, we consider the case where the parameters are set to $\phi = 0$ and $\Delta = 1$, and the elements of the optimal POVM are

\begin{equation}
    \Pi_1 =
\begin{bmatrix}
  0.3791 + 0.0000i & 0.2140 + 0.0023i & 0.2140 + 0.0023i & 0.3791 + 0.0081i \\
0.2140 - 0.0023i & 0.1209 + 0.0000i & 0.1209 - 0.0000i & 0.2140 + 0.0023i \\
0.2140 - 0.0023i & 0.1209 + 0.0000i & 0.1209 + 0.0000i & 0.2140 + 0.0023i \\
0.3791 - 0.0081i & 0.2140 - 0.0023i & 0.2140 - 0.0023i & 0.3791 + 0.0000i
\end{bmatrix},
\end{equation}

\begin{equation}
    \Pi_2 =
\begin{bmatrix}
  0.3111 + 0.0000i & -0.1043 - 0.2188i & -0.1043 - 0.2188i & -0.1959 + 0.2417i \\
-0.1043 + 0.2188i &  0.1889 + 0.0000i &  0.1889 + 0.0000i & -0.1043 - 0.2188i \\
-0.1043 + 0.2188i &  0.1889 + 0.0000i &  0.1889 + 0.0000i & -0.1043 - 0.2188i \\
-0.1959 - 0.2417i & -0.1043 + 0.2188i & -0.1043 + 0.2188i &  0.3111 + 0.0000i
\end{bmatrix},
\end{equation}
\begin{equation}
    \Pi_3 =
\begin{bmatrix}
  0 & 0 & 0 & 0 \\
  0 & 0.5000 & -0.5000 & 0 \\
  0 & -0.5000 & 0.5000 & 0 \\
 0 & 0 & 0 & 0
\end{bmatrix},
\end{equation}
and $\Pi_4=\mathbb{I}-\Pi_1-\Pi_2-\Pi_3$, with basis $\{|00\rangle, |01\rangle, |10\rangle, |11\rangle\}$. We also find the results show rank$(\Pi_k)=1,\ \forall k=1,2,3,4$ and $\mathrm{Tr}(\Pi_m\Pi_n)\ne 0,$ $\forall m\ne n$. It is not a projective measurement but a rank-one POVM. The quantity $\mathrm{Tr}(\mathbf{Q}_2^{-1}\mathbf{F}_2)$ of this case is about 1.36, slightly smaller than 1.5. Besides, Chen et. al. proposed tighter trade-off relations for the precision of multiple parameters which can be formulated as semi-definite programming \cite{RN42}. By using equation (26) from that work, the optimal value of $\mathrm{Tr}(\mathbf{Q}_2^{-1}\mathbf{F}_2)$ is 1.36, which confirms that the measurement mentioned above is optimal.
\medskip

\textbf{Acknowledgments}

This work was supported by lnnovation Program for Quantum Science and Technology (Grant No. 2024ZD0300900), the National Natural Science Foundation of China (Grant Nos. 12347104, U24A2017, 12461160276, and 12504418), the National Key Research and Development Program of China (Grant No. 2023YFC2205802), and the Natural Science Foundation of Jiangsu Province (Grant Nos. BK20243060 and BK20233001).

\textbf{Conflicts of Interest}

The authors declare no conflicts of interest.
\medskip

%

\bibliography{reference.bib}

\end{document}